\documentclass[aps,prl,twocolumn,superscriptaddress]{revtex4}

\usepackage{graphicx}
\usepackage{dcolumn}
\usepackage{bm}
\usepackage{color}

\begin{document}
\title{Quantisation of Hopping Magnetoresistance Prefactor in Strongly Correlated Two-Dimensional Electron Systems}
\author{Matthias Baenninger}
\email{matthias.baenninger@cantab.net}
\affiliation{Cavendish Laboratory, University
of Cambridge, J.J. Thomson Avenue, Cambridge CB3 0HE, United
Kingdom.}
\author{Arindam Ghosh}
\affiliation{Department of
Physics, Indian Institute of Science, Bangalore 560 012, India.}
\author{Michael Pepper}
\affiliation{Cavendish Laboratory, University of Cambridge, J.J.
Thomson Avenue, Cambridge CB3 0HE, United Kingdom.}
\author{Harvey E. Beere}
\affiliation{Cavendish Laboratory, University of Cambridge, J.J.
Thomson Avenue, Cambridge CB3 0HE, United Kingdom.}
\author{Ian Farrer}
\affiliation{Cavendish Laboratory, University of Cambridge, J.J.
Thomson Avenue, Cambridge CB3 0HE, United Kingdom.}
\author{Paula Atkinson}
\affiliation{Cavendish Laboratory, University of Cambridge, J.J.
Thomson Avenue, Cambridge CB3 0HE, United Kingdom.}
\author{David A. Ritchie}
\affiliation{Cavendish Laboratory, University of Cambridge, J.J.
Thomson Avenue, Cambridge CB3 0HE, United Kingdom.}
\date{\today}
\maketitle

{\bf We report an universal behaviour of hopping transport in strongly interacting mesoscopic two-dimensional electron systems (2DES). In a certain window of background disorder, the resistivity at low perpendicular magnetic fields follows the expected relation $\rho(B_\perp) = \rho_{\rm{B}}\exp(\alpha B_\perp^2)$.
The prefactor $\rho_{\rm{B}}$ decreases exponentially with increasing electron density but saturates to a finite value at higher densities. Strikingly, this value is found to be universal when expressed in terms of absolute resistance and and shows quantisation at  $R_{\rm{B}}\approx h/e^2$  and $R_{\rm{B}}\approx 1/2$ $ h/e^2$. We suggest a strongly correlated electronic phase as a possible explanation.}

\section{Introduction} Transport in 2DES in a regime where both disorder and electron-electron interactions are strong is still not well understood. Mesoscopic 2DES in modulation doped GaAs/AlGaAs heterojunctions are an ideal system for the investigation of this regime since they allow for a control of disorder strength by varying the spacer width $\delta_{\rm{sp}}$ between 2DES and the charged dopants, while largely overcoming the problem of long-range charge inhomogeneities that often dominate transport on a macroscopic length scale~\cite{Ghosh2004,Baenninger2005}. Here, we present results of extensive magnetoresistance (MR) measurements in hopping transport in mesoscopic 2DES of varying disorder strength. Sample details and experimental techniques are described in Refs.~\cite{Ghosh2004,Baenninger2005}.
\section{Results and Discussion}
In the strongly localised regime of 2DES, transport is assumed to occur by phonon assisted tunnelling of electrons between trap sites arising from the background disorder potential. The compression of the wave function in a weak perpendicular magnetic field leads to a MR $\rho(T, B)=\rho_{\rm{B}}(T)\exp(\alpha B^{2})$~\cite{Shklovskii1983}. Recently, investigations of the exponent $\alpha$ in mesoscopic 2DES in modulation-doped GaAs/AlGaAs heterojunctions revealed an universal behaviour~\cite{Ghosh2004,Baenninger2005}: The average hopping distance $r_{\rm{ij}}$ appeared to be determined only by the electron density $n_{\rm{s}}$ and, assuming a localisation length $\xi=a_{\rm{B}}^{*}$ (the effective Bohr radius), it was found to follow the relation $r_{\rm{ij}}\approx r_{\rm{ee}}\cong 1/\sqrt{n_{\rm{s}}}$ ($r_{\rm{ee}}$ the average electron-electron separation) independent of background disorder. An alternative interpretation of these results has been suggested in form of tunnelling between electron droplets~\cite{Tripathi2006}.

\begin{figure*}[t]
\centering
\includegraphics[width=1\textwidth]{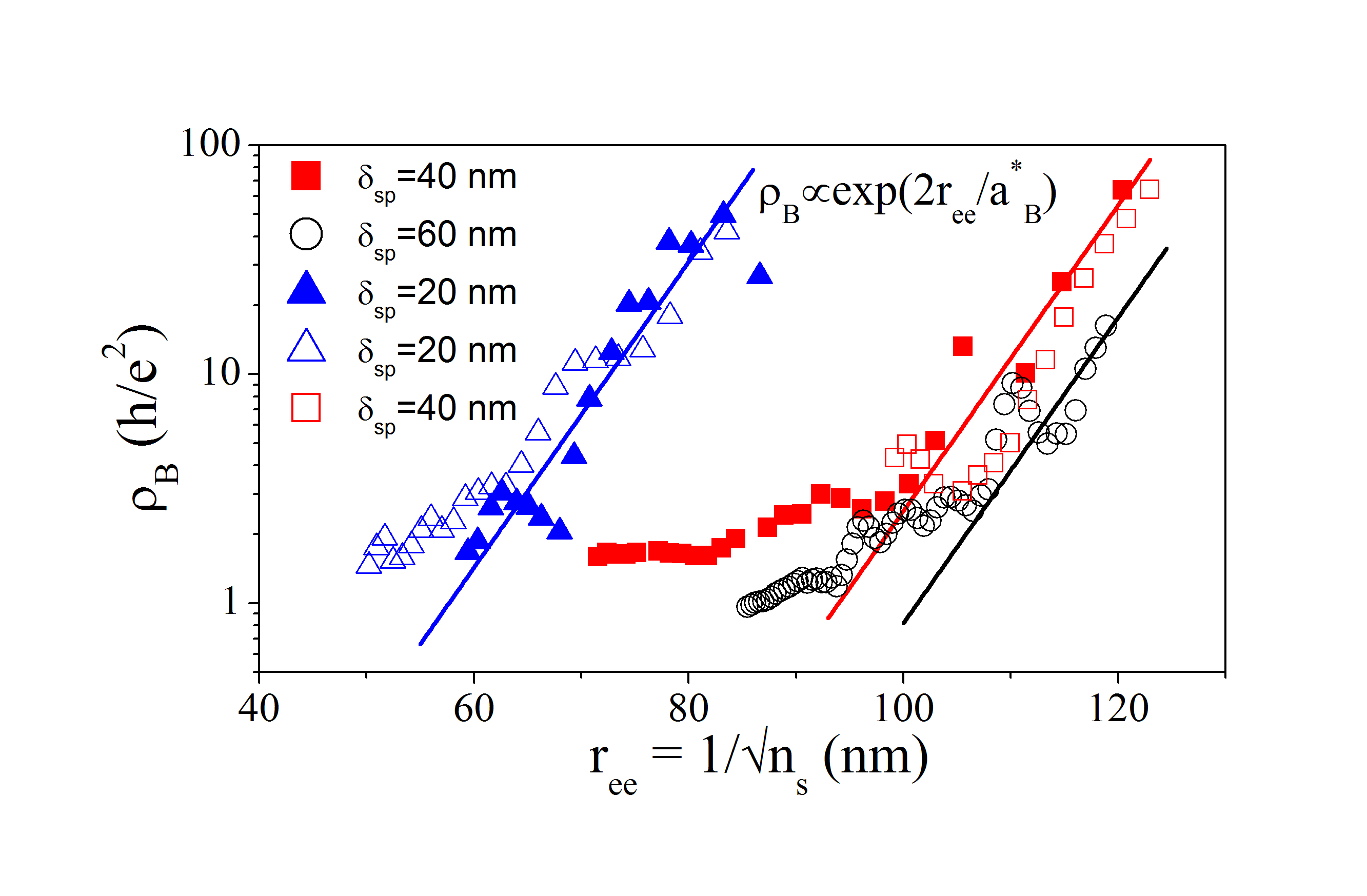}
\caption{$\rho_{\rm{B}}$ as a function of $r_{\rm{ee}}$, showing exponential and saturated behaviour ($T$=300 mK).} 
\label{F_MR}\end{figure*}

\begin{figure*}[t]
\centering
\includegraphics[width=1\textwidth]{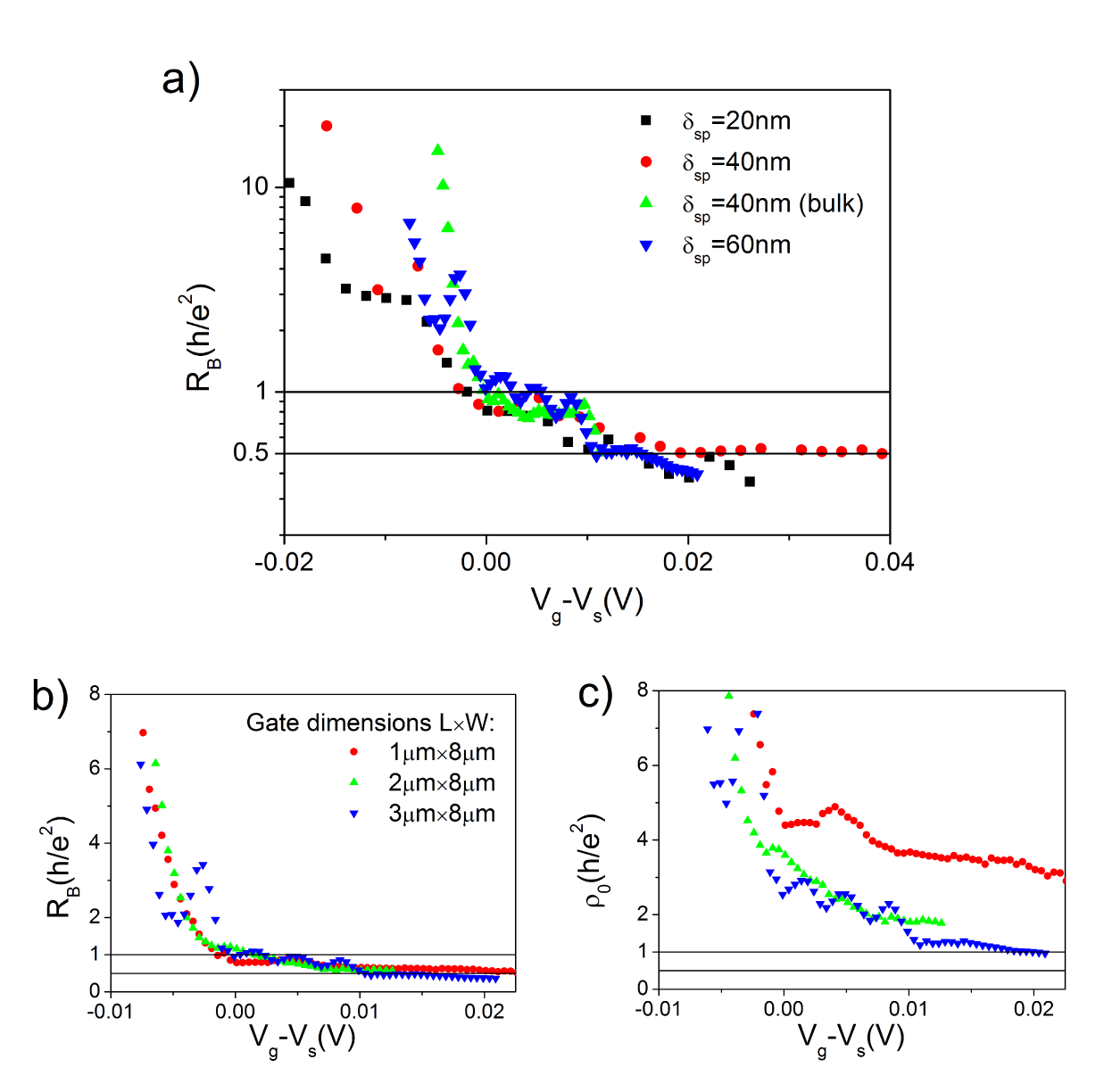}
\caption{a) MR prefactor in terms of absolute resistance device from four different wafers. They all show plateaux at $R_{\rm{B}}\approx h/e^2$ and $h/2e^2$.\newline
b) and c) Comparison of $\rho_{\rm{B}}$ and $R_{\rm{B}}$ for devices with varying dimensions, confirming an universality in resistance rather than resistivity. Gate voltages have been shifted by a value of $V_{\rm{S}}$ for better representation ($T$=300 mK).} 
\label{F_RBRhoB}\end{figure*}

\begin{figure*}[t]
\centering
\includegraphics[width=1\textwidth]{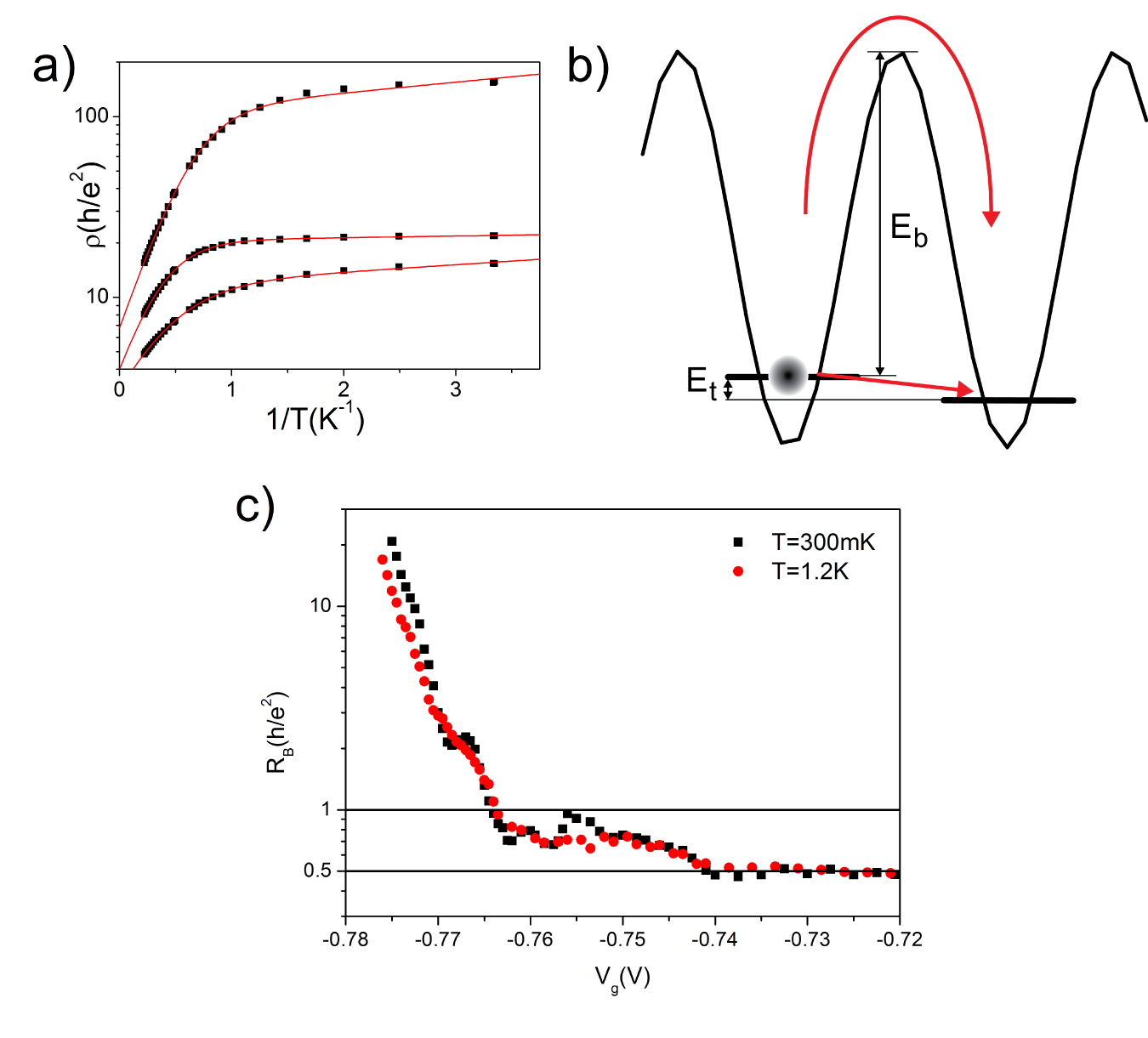}
\caption{ a) Transition between two regimes of temperature dependence in the resistance of mesoscopic 2DES. b) Schematic of hopping transport and energy scales involved. c) $R_{\rm{B}}$ at 0.3 K and 1.2 K for the same device. Results are almost identical for both temperatures.} 
\label{F_TDep}\end{figure*}

In this article, we focus on the behaviour of the prefactor $\rho_{\rm{B}}$ and show that it confirms the findings described above, but also raises further questions. $\rho_{\rm{B}}$ was determined by fitting the MR expression given above to the data (see Refs.~\cite{Ghosh2004,Baenninger2005} for details).
In case of nearest-neighbour hopping, the prefactor behaves as~\cite{Shklovskii1983}
\begin{equation}\label{E_RhoB}
\rho_{\rm{B}}=\rho_{0}\exp(E_{\rm{t}}/k_{\rm{B}}T)\exp(2r_{\rm{ij}}/\xi).
\end{equation}
This relation allows for an independent verification of the findings $\xi=a_{\rm{B}}^{*}$ and $r_{\rm{ij}}\approx r_{\rm{ee}}$. Indeed, as Fig.~\ref{F_MR} shows, the expression $\rho_{\rm{B}}\propto \exp(2r_{\rm{ee}}/a_{\rm{B}}^{*})$ agrees with the experimental data at low electron densities.
However, at higher $n_{\rm{s}}$, a deviation from this exponential behaviour is observed for all devices with a saturation of $\rho_{\rm{B}}$ at a value of order $h/e^2$. In fact, the saturation value turns out to be universal for all devices when expressed in terms of absolute resistance $R=\rho\times L/W$ ($L$ the length, $W$ the width of the gate) and form plateaux at $R_{\rm{B}}\approx h/e^2$ and $h/2e^2$. This is shown in Fig.~\ref{F_RBRhoB} a), for devices from four different wafers with spacer width varying from 20 nm to 60 nm. Figs.~\ref{F_RBRhoB} b) and c) directly compare a set of three devices from the same wafer but with different gate dimensions and, hence, ratios between $\rho_{\rm{B}}$ and $R_{\rm{B}}$. They clearly confirm that the saturation is, indeed, universal in resistance rather than resistivity.
Eq.~\ref{E_RhoB} predicts an exponential temperature dependence of $\rho_{\rm{B}}$ and, hence, $R_{\rm{B}}$. However, we observe a virtually temperature independent prefactor from $T$=300 mK to 1.2 K, as shown in Fig.~\ref{F_TDep} c). This is related to an unusual temperature dependence of the resistance itself that has been observed in the localised regime of mesoscopic devices~\cite{Ghosh2004,Baenninger2005,Baenninger2007}: After an activated part at higher temperature, a strong weakening of the temperature dependence was seen at $T\lesssim 1$ K (see Fig.~\ref{F_TDep} a). A possible explanation for this behaviour is that of a two component transport of the form $\rho(T)^{-1} = \rho_{\rm{b}}^{-1}\exp(-E_{\rm{b}}/k_{\rm{B}}T) + \rho_{\rm{t}}^{-1}\exp(-E_{\rm{t}}/k_{\rm{B}}T)$, with the first term representing activated hopping across the potential barriers between electron trap sites and the second term being thermally assisted tunnelling between nearest neighbour states (see schematic Fig.~\ref{F_TDep} b). Under the condition $E_{\rm{b}}\gg E_{\rm{t}}$ and $\rho_{\rm{b}}\ll \rho_{\rm{t}}$, the first term will dominate at higher temperature and the second one at low temperatures. Indeed, the expression above can be fitted (solid lines) very well to our data (symbols) as Fig.~\ref{F_TDep} a) shows, with $\rho_{\rm{b}}^{-1}\exp(-E_{\rm{b}}/k_{\rm{B}}T) \ll \rho_{\rm{t}}^{-1}\exp(-E_{\rm{t}}/k_{\rm{B}}T)$ at lowest $T$. The fitting also reveals $E_{\rm{t}}\ll k_{\rm{B}}T$ in the whole range of accessible temperature. This means that Eq.~\ref{E_RhoB} becomes $\rho_{\rm{B}} \approx \rho_{0}\exp(2r_{\rm{ij}}/\xi)$, which explains the temperature independence of $\rho_{\rm{B}}$.

We suggest a transport mechanism mediated by defects such as interstitials or vacancies in an interaction-induced, disorder-stabilised electron solid as a possible explanation for the observations described above. It provides a natural explanation for the observation $r_{\rm{ij}}\approx r_{\rm{ee}}$~\cite{Ghosh2004,Baenninger2005}, but may also explain the the unusual temperature dependence~\cite{Baenninger2007} and the behaviour of $R_{\rm{B}}$. In a picture of hopping between trap sites caused by random disorder, one would expect $E_{\rm{t}}$ and $E_{\rm{b}}$ to be of the same order of magnitude. In an electron solid, the trap sites are formed by the electrons themselves. This allows for a much greater regularity and explains the smallness of $E_{\rm{t}}$, which would be zero in a perfect electron crystal, but becomes finite due to the presence of finite disorder.

For the prefactor of the exponential temperature dependence in hopping transport, a value of $\approx h/e^2$ has been observed experimentally and interpreted as possible evidence of an electron-electron interaction mediated energy transfer mechanism~\cite{Khondaker1999,Mason1995,Aleiner1994a}. In our case it may well be that the hopping is assisted by phonons in the electron solid itself, rather than in the GaAs crystal. The universality and quantisation at  $R_{\rm{B}}\approx h/e^2$  and $ h/2e^2$ in absolute resistance could be explained in terms of a correlated hopping or possibly by a dominance of transport by a one dimensional path with highest conductance between source and drain, which could be caused by disorder or an interaction driven stripe formation in the 2DES~\cite{Koulakov1996}. However, details of such mechanisms are not understood at present.
\section{Conclusions}
We have demonstrated an universal behaviour of hopping conductance in 2DES with the magnetoresistance prefactor being quantised at the values $R_{\rm{B}}\approx h/e^2$  and $h/2e^2$. Supported by earlier results, the observation has been interpreted as possible evidence of defect transport in a pinned electron solid, where the energy transfer is mediated by lattice vibrations of the electron solid itself.
\section*{Acknowledgements}
MB acknowledges the support of RTN COLLECT, the Sunburst Trust and the Cambridge Overseas Trust.
We thank M. Mueller for a valuable discussion.

\begin{thebibliography}{1}

\bibitem{Ghosh2004}
A.~Ghosh, M.~Pepper, H.~E. Beere, and D.~A. Ritchie,
\newblock Phys. Rev. B 70 (2004) 233309. 

\bibitem{Baenninger2005}
M.~Baenninger, A.~Ghosh, M.~Pepper, H.~E. Beere, I.~Farrer, P.~Atkinson, and
  D.~A. Ritchie,
\newblock Phys. Rev. B 72 (2005) 241311.

\bibitem{Shklovskii1983}
B.~I. Shklovskii,
\newblock Sov. Phys. Semicond. 17 (1983) 1311.

\bibitem{Tripathi2006}
V.~Tripathi, and M.~P.~Kennett,
\newblock Phys. Rev. B 74 (2006) 195334.

\bibitem{Baenninger2007}
M.~Baenninger, A.~Ghosh, M.~Pepper, H.~E. Beere, I.~Farrer, and D.~A. Ritchie,
\newblock Submitted (2007).

\bibitem{Khondaker1999}
S.~I. Khondaker, I.~S. Shlimak, J.~T. Nicholls, M.~Pepper, and D.~A. Ritchie,
\newblock Phys. Rev. B 59 (1999) 4580.

\bibitem{Mason1995}
W.~Mason, S.~V. Kravchenko, G.~E. Bowker, and J.~E. Furneaux,
\newblock Phys. Rev. B 52 (1995) 7857.

\bibitem{Aleiner1994a}
I.~L. Aleiner, D.~G. Polyakov, and B.~I. Shklovskii,
\newblock Proc. 22nd Int. Conf. Phys. Semicond., Vancouver, 1994,
  1 (1994) 787.

\bibitem{Koulakov1996}
A.~A. Koulakov, M.~M. Fogler, and B.~I. Shklovskii,
\newblock Phys. Rev. Lett. 76 (1996) 499.

\end{thebibliography}
\end{document}